\documentclass[%
 aps,
 sd,%
 amsmath,amssymb,
reprint,%
nofootinbib
]{revtex4-1}

\usepackage{color}
\usepackage{graphicx}	
\usepackage{dcolumn}
\usepackage{bm}

\newcommand\simlt{\lower.5ex\hbox{$\; \buildrel < \over \sim \;$}}
\newcommand\simgt{\lower.5ex\hbox{$\; \buildrel > \over \sim \;$}}

\def\apjl{ApJL }

\def\apj{Astropys. J. }

\def\apjs{ApJS }

\def\mnras{Mon. Not. Roy. Astron. Soc. }

\def\ssr{Space Sci. Rev. }

\def\prl{Phys. Rev. Lett. }
\def\physrep{Physics Reports}

\begin{document}

\title{Plasma kinetic effects in relativistic radiation mediated shocks}
\author{Amir Levinson }
\affiliation{School of Physics \& Astronomy, Tel Aviv University, Tel Aviv 69978, Israel}

\date{\today}
\begin{abstract}
Fast shocks that form in optically thick media are mediated by Compton scattering and, if relativistic, pair creation.
Since the radiation force acts primarily on electrons and positrons, the question arises of how the force is mediated 
to the ions which are the dominant carriers of the shock energy.    It has been widely thought that a small charge separation induced
by the radiation force generates electric field inside the shock that decelerates the ions.   In this paper we argue that,
while this is true in sub-relativistic shocks which are devoid of positrons,
in relativistic radiation mediated shocks (RRMS), which are dominated by newly created  $e^+e^-$ pairs, additional coupling
is needed, owing to the opposite electric force acting on electrons and positrons.
Specifically, we show that dissipation of the ions energy must involve collective plasma interactions.  
 By constructing a multi-fluid model for RRMS that incorporates friction forces, we estimate that 
momentum transfer between electrons and positrons (and/or ions) via collective interactions on scales of tens to thousands proton skin depths, 
depending on whether friction is effective only between $e^+e^-$ pairs or also between pairs and ions,
is sufficient to coupe all particles and radiation inside
the shock into a single fluid.   This leaves open the question whether in relativistic RMS 
particles can effectively accelerate to high energies by scattering off plasma turbulence.    
Such acceleration might have important consequences for relativistic shock breakout signals.
\end{abstract} 


\maketitle

\section{Introduction}
Radiation mediated shocks (RMS) play a key role in a plethora of extreme cosmic phenomena.  The early emission from 
shock breakout in different types of supernovae, prompt photospheric emission in GRBs, and the gamma ray emission 
that accompanied the gravitational wave signal in neutron star mergers are some notable examples (for a recent review 
see Levinson and Nakar \cite{levinson2020} and references therein).   
In marked difference to collisionless shocks, that form in optically thin media and in which dissipation is mediated 
by collective plasma processes on skin depth scales \cite{spitkovsky2008a,spitkovsky2008b,lemoine2010,shaisultanov2012,derishev2016,lemoine2019}, 
RMS, that form in optically thick media,  are mediated by 
Compton scattering and, under certain conditions, pair creation, on scales of the order of
the Thomson length (i.e., the scattering mean free path).   There exists another type of relativistic RMS, that form in optically thin media \cite{derishev2016b};
those are radiation mediated but not radiation dominated, and involve a different microphysics.  Our interest here focuses on radiation dominated RMS.

Much effort has been devoted in the last decade to developing analytical \cite{levinson2008,katz2010,nakar2010,bromberg2011,sapir2011,nakar2012,sapir2013,
beloborodov2017b,granot2018,ioka2018,lundman2018a,lundman2018b,lyutikov2018,derishev2018}
and numerical \cite{budnik2010,ito2018a,ito2020,beloborodov2017a} models of 
RMS in the Newtonian and relativistic regimes.    These studies indicate that there are essentially three 
important regimes which are distinguished by the shock velocity $\beta_u$, henceforth measured in units
of $c$ (see Ref. \cite{levinson2020} for a detailed account):  Slow shocks  ($\beta_u \simlt 0.05$) in which 
the radiation is in thermodynamic equilibrium
and the breakout temperature depends rather weakly on the shock velocity; Fast Newtonian 
shocks ($0.05\simlt \beta_u \simlt 0.5$), in which the radiation is out of thermodynamic equilibrium 
and the temperature is determined by the amount of photons produced in the immediate downstream;
Relativistic shocks ($\beta_u\gamma_u >0.5$), for which the shock structure and
emission are strongly affected by vigorous pair creation.   Those relativistic RMS are at the main focus of this paper. 

All RMS models tacitly assume that the different plasma species (ions, electrons, positrons and photons) are tightly coupled,
and invoke the single fluid  approximation (in the sense that all species share the same local center of momentum frame).
Since the cross section for Compton scattering off protons is smaller by a factor $(m_p/m_e)^2$ than that for electrons
(and positrons when present), 
the radiation force acting on the ions is completely negligible.   This raises the question of how the radiation force is 
mediated to the ions.  The conventional wisdom has been that a tiny charge separation, induced by the radiation force experienced
by the charged leptons, generates electrostatic field that decelerates the ions.  We shall show that this notion,
which seems rather trivial,  holds true only in cases where the plasma flowing through the shock is devoid of positrons, albeit 
as pointed out in Ref. \cite{derishev2018} the presence of 
ions with different charge-to-mass ratio can lead to a considerable Ohmic heating that may alter the shock microphysics.   
We will not consider this complication in what follows, and assume for simplicity that the shock propagates in a pure hydrogen gas. 
Under this condition we show, in Sec. \ref {sec:subR}, that in subrelativistic RMS, where pair creation is slow and the positron density
is vanishingly small, a tiny difference in the velocities of the proton and electron fluids, roughly 
$\Delta \beta/\beta = (l_p/L)^2\simlt 10^{-18}$,  is sufficient to generate the required electric field that decelerates 
the protons, where  $l_p=c/\omega_P$ is the proton skin depth defined in Eq. (\ref{eq:skindepth}) below and $L$ the width of 
the shock transition layer.   The same electric field exerts a force opposite to the radiation force on the 
electrons, nearly cancelling it out.  The residual force, roughly a factor $m_e/m_p$ smaller than the radiation force,
is exactly what needed to decelerate the electrons at the same rate as the protons.
 
 The situation is drastically different in relativistic RMS, in which $e^+e^-$ pairs are overabundant, owing to the asymmetric 
 net force acting on the pairs system.   The salient point is that the electric field required to decelerate the protons 
 exerts opposite forces on electrons and positrons.     As shown below, once the density of newly created positrons 
 approaches the baryon density, which in relativistic RMS occurs at the onset of the shock transition layer,  
 the charge density, and ultimately the electric field, reverse sign.
 This leads to decoupling of the different species early on.   More precisely, while the pairs decelerate 
 by the radiation drag force, the protons continue to propagate undisturbed  at their initial velocity. 
 In practice this should lead to a rapid  growth (on skin depth scales) of plasma instabilities, as seen in plasma
 simulations of collisionless shocks \cite{spitkovsky2008a,lemoine2019}, 
 that should provide  tight couplings between the various charged species.   Self-consistent calculations of relativistic RMS that 
 take into account plasma kinetic effects are currently infeasible, as they require a huge dynamic range, from the skin 
 depth 
 \begin{equation}
l_p = \frac{c}{\omega_P} \approx 0.5 \gamma_u^{1/2}\left(\frac{n_u}{10^{15}\, cm^{-3}}\right)^{-1/2} \quad \rm cm, \label{eq:skindepth}
\end{equation}
 to the Thomson length
 \begin{equation}
\lambda= (\sigma_T n_u\gamma_u)^{-1} \approx 10^9 \gamma_u^{-1}\left(\frac{n_u}{10^{15}\, cm^{-3}}\right)^{-1} \quad \rm cm, \label{eq:lambda}
\end{equation}
here $\omega_p=(4\pi e^2 n_u/m_p\gamma_u)^{1/2}$ is the (proton) plasma frequency, $n_u$ the proper density far upstream of the shock 
and $\gamma_u$ the shock Lorentz factor.

In this paper we construct a multi-fluid model for relativistic RMS that incorporates electromagnetic forces, as well as  phenomenological 
friction force terms that represent the effect of collective plasma interactions.   Our model generalizes the single fluid 
model presented in Ref. \cite{granot2018}.   We show that relativistic RMS solutions exist provided that collective plasma 
interactions can give rise to a significant momentum transfer between electrons and positrons on  scales of tens to 
hundreds $l_p$. If the friction between protons and pairs is also strong then an even weaker coupling (larger 
interaction length), by a factor of several hundreds, is sufficient.    

The anticipated growth of turbulence inside the shock revives the issue of particle acceleration in RMS.   The claim 
that particle acceleration is prohibited in RMS by the vast separation of kinetic and radiation 
scales (e.g., Ref. \cite{levinson2008}) may not hold in the relativistic regime if strong turbulence is indeed generated, as argued here.  
Second order Fermi acceleration by magnetic turbulence may turn out to be effective.   If this will be corroborated by detailed 
simulations it can have important implications for the high-energy emission from relativistic shock breakouts.

\section{basic equations}
The fluid inside a relativistic RMS is a mixture of protons, positrons electrons and radiation,  
with proper densities $n_p, n_+, n_-, n_r$, respectively, moving at 4-velocities $u_p^\mu, u_+^\mu, u_-^\mu, u_r^\mu$, with
the notation $u^\mu = (\gamma, \gamma {\bm \beta})$ henceforth adopted.   The energy-momentum tensor of
each charged species can be expressed as
\begin{equation}
T_a^{\mu\nu}= m_ac^2 n_a h_a u_a^\mu u_a^\nu +g^{\mu\nu} p_a,
\end{equation}
with the index $a$ running over  $(p, +, -)$. Here $m_a$ is the particle mass, $p_a$ is the pressure of species $a$ 
and $h_a$ the corresponding dimensionless enthalpy per particle.    A similar expression can be derived for the 
energy-momentum tensor of the radiation, $T_r^{\mu\nu}$, with $m_\alpha c^2 h_\alpha = 4kT_r$, here $T_r$ being the 
radiation temperature.
Quite generally, the shock temperature is well below the proton mass, so that  to a good approximation $h_p=1$ can be adopted.
For the leptons a relativistic equation of state, $h_\pm = 4p_\pm/m_ec^2n_\pm = 4\hat{T}_\pm$, 
is a good description in the relativistic regime, whereas in subrelativistic shocks, where only electrons are present,  
$h =1+ 5\hat{T}/2$.   Here $\hat{T} = kT /m_ec^2 $ denotes the temperature in units of the electron mass.

Conservation of baryons and quanta number read:
\begin{eqnarray}
\partial_\mu (n_pu_p^\mu)=0, \label{eq:mass_p}\\ 
\partial_\mu (n_\pm u_\pm^\mu)= q_\pm = q/2, \label{eq:mass_pairs} \\
\partial_\mu (n_r u_r^\mu) = -q,
\end{eqnarray}
where the source terms $q_+ = q_-=q/2$ account for pair production.  The dynamics of the system is governed by the equations
\begin{eqnarray}
\partial_\mu T_{p}^{\mu\nu}= f_p^\nu +g_p^\nu,\\ 
\partial_\mu T_{\pm}^{\mu\nu}= S_\pm^\nu + f_\pm^\nu  +g_\pm^\nu,\\  
\partial_\mu T_{r}^{\mu\nu}=- (S_+^\nu + S_-^\nu),
\end{eqnarray}
where $S_\pm^\mu$ accounts for energy and momentum exchange between pairs and photons, 
\begin{eqnarray}
g_p^\nu = -m_ec \chi_{pe} [n_pn_+ (u_p^\nu-u_+^\nu) + n_pn_- (u_p^\nu-u_-^\nu)],
\end{eqnarray}
and 
\begin{eqnarray}
\begin{split}
g_\pm^\nu = & \pm m_ec \chi_{ee} n_+n_-(u^\nu_- - u_+^\nu) \\
&+ m_ec^2\chi_{pe} n_pn_\pm (u_p^\nu-u_\pm^\nu),
\end{split}
\end{eqnarray}
represent internal friction between protons and pairs, and between electrons and positrons, respectively, with
$\chi_{pe}$ and $\chi_{ee}$ being the corresponding dynamical coefficients\footnote{The units of the dynamical 
coefficients adopted here are different than the standard choice, e.g., Ref. \cite{zenitani2009}. The two 
are related through $\tau_{fr\alpha}=m_e\chi_{\alpha}$}  (for simplicity 
we assume the same coefficient for the interaction of protons with electrons and positrons), 
and $f_a^\mu$ the electromagnetic force density acting on the charged fluids, given explicitly by
\begin{equation}
f^\mu_{a}=q_\alpha n_\alpha F^{\mu\nu}u_{\alpha \nu}= q_\alpha n_\alpha \gamma_\alpha (\bm \beta_\alpha\cdot \bm E, \bm E+ \bm \beta_\alpha \times \bm B)
\end{equation}
for $\alpha = (p,+,-)$, where $F^{\mu\nu}$ is the electromagnetic tensor, $\bm E, \bm B$ 
denote the electric and magnetic fields measured in some global inertial frame to be specified later on, and 
$q_p=q_+ =e$, $q_-= -e$ are the electric charges of the designated species. 
The electromagnetic field satisfies Maxwell's equations:
\begin{eqnarray}
\partial_\mu F^{\mu\nu}= \frac{4\pi}{c} j^\mu,\\
\partial_\mu \left(\frac{1}{2}\epsilon^{\mu\nu\alpha\beta} F_{\alpha\beta}\right)=0,
\end{eqnarray}
with the electric 4-current given explicitly by
\begin{equation}
j^\mu = ec (n_pu_p^\mu + n_+ u_+^\mu -n_- u_-^\mu).
\end{equation}
Equations (\ref{eq:mass_p}) and (\ref{eq:mass_pairs}) combined yield the charge conservation condition, $\partial_\mu j^\mu=0$, as required.

\section{A planar shock model}
Consider a steady planar shock propagating in the $x$ direction into a cold, unmagntized medium.  
In the static shock frame far upstream is located at $x=-\infty$,
and all fluid quantities depend on the coordinate $x$ only.    

The boundary conditions far upstream (formally at $x=-\infty$) are:
\begin{eqnarray}
n_p=n_- = n_u,\quad n_+=0, \label{eq:bc_n}\\
u_p^\mu = u_-^\mu= \gamma_u (1,\beta_u,0,0), \label{eq:bc_u}\\
\bm B =\bm E=0. \label{eq:bc_EB}
\end{eqnarray}
The fields $\bm E$ and $\bm B$ and the 4-velocities are henceforth measured in the rest frame of the shock.

 Equation (\ref{eq:mass_p}) implies that the baryon flux through the shock, $J_p = n_pu_p^x$, is constant.
Likewise, charge conservation, $\partial_\mu j^\mu=0$, yields $\partial_x j^x=0$.  
Applying the boundary conditions (\ref{eq:bc_n}) and (\ref{eq:bc_u}) we then find that the $x$ 
component of the electric current density vanishes:
\begin{equation}
j^x = ec(J_p +n_+u^x_+ -n_- u^x_-)=0.
\end{equation}
From $\nabla\cdot \bm B=0$ and the boundary condition (\ref{eq:bc_EB}) we obtain $B^x=0$.    
Choosing the coordinate system such that $\bm B = B\hat{y}$ and using  Amper's law, $\nabla \times {\bf B}=4\pi {\bf j}/c$, 
yields,
\begin{equation}
\partial_x B = \frac{4\pi}{c} j^z = 4\pi e (n_pu_p^z +n_+u^z_+ -n_- u^z_-)\label{eq:Amper}.
\end{equation}
Faraday's law, $\nabla \times {\bf E}= 0 $, reduces to $\partial_x E_z= \partial_x E_y=0$, and applying the far upstream conditions 
we have $E_z=E_y=0$, whereby ${\bm E}= E \hat{x}$.   Gauss' law then reduces to
\begin{equation}
\partial_x E = 4\pi e(n_p \gamma_p + n_+\gamma_+ - n_- \gamma_-).
\end{equation}
Note that $\bm j \cdot \bm E=0$, so no Ohmic heating is expected in this case. This will no longer be true if the shock develops
electrostatic oscillations, $\partial_t \bm E= -4\pi \bm j$.

We now turn to consider the momentum equations.   In the transverse direction the system is supposed to be uniform, hence
$S^y_\pm=S^z_\pm=0$.  The transverse momentum equations then simplify to:
\begin{eqnarray}
\partial_x u_p^z = \frac{e }{m_pc^2} B +\frac{g_p^z}{m_pc^2J_p},\label{eq:mom_trans_p}\\
\partial_x(m_ec^2n_\pm h_\pm u_\pm^x u_\pm^z) = e n_\pm  u_\pm^x B +g_\pm^z.\label{eq:mom_trans}
\end{eqnarray}
Clearly, the only solution to  Eqs.(\ref{eq:Amper}), (\ref{eq:mom_trans_p}) and (\ref{eq:mom_trans}) that satisfies the far upstream conditions (\ref{eq:bc_n})-(\ref{eq:bc_EB}) is $B=u_p^z=u_\pm^z=0$.   It is now seen that a stationary, planar RMS propagating in an unmagnetized
medium can only generate electrostatic field ($\bm B=0$).

In the longitudinal direction (along $x$) we have 
\begin{eqnarray}
\partial_x (m_pc^2J_pu_p^x) +\partial_x p_p= e n_p\gamma_pE +g_p^x,\\ 
\partial_x(m_ec^2n_\pm h_\pm u_\pm^x u_\pm^x) +\partial_x p_\pm = \\ \nonumber
 \pm e n_\pm\gamma_\pm E + S^x_\pm +g_\pm^x, \label{eq:mom_pairs} \\
\partial_x T_r^{xx} = -(S^x_+ + S^x_-).
\end{eqnarray}

Finally, the energy equations read:
\begin{eqnarray}
\partial_x \gamma_p = \frac{e }{m_pc^2} E + \frac{g_p^0}{m_pc^2J_p} , \\
\partial_x(m_ec^2n_\pm h_\pm \gamma_\pm u_\pm^x) = \pm  e n_\pm u^x_\pm E  \\ \nonumber
+ S^0_\pm +  g_\pm^0,\label{eq:energ_pairs}\\ 
\partial_x T_r^{x0} = -(S^0_+ + S^0_-).
\end{eqnarray}

The above set of equations will be solved below in the subrelativistic and highly relativistic limits under certain approximations.

\begin{figure}
\centering
\centerline{ \includegraphics[width=8cm]{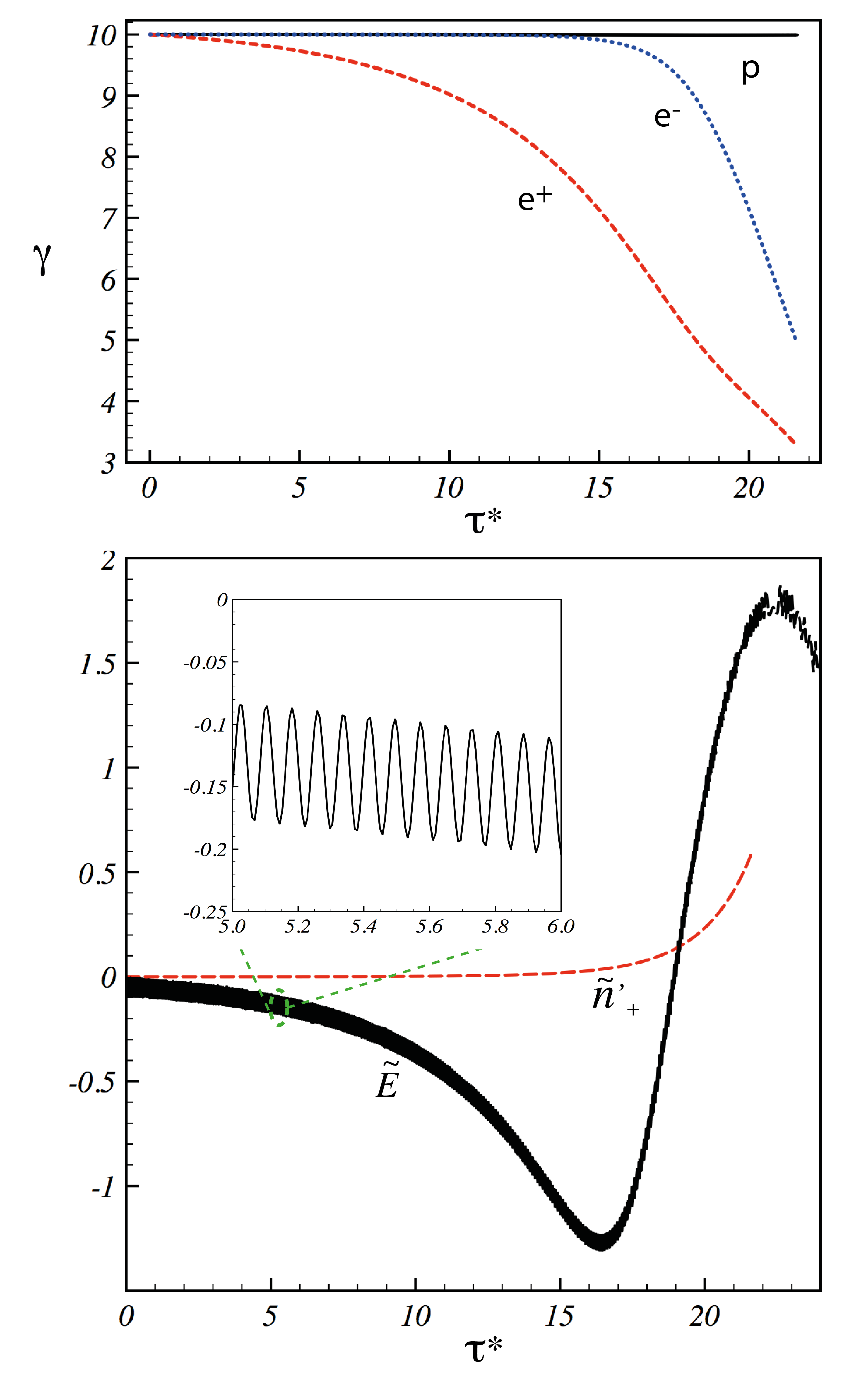}}  
\caption{Solutions of the shock equations for frictionless fluids, $\chi_{pe}=\chi_{ee}=0$.  The upper panel shows profiles of the Lorentz 
factors of protons (p), electrons ($e^-$) and positrons ($e^+$), plotted as functions of the pair-loaded optical depth $\tau^*$, 
and the lower panel displays the dimensionless electric field $\tilde{E}$ (black line)
and positron density $\tilde{n}'_+$ (red dashed line).   The inset shows a zoom of the electric field solution. }
\label{fig:no_fric}
\end{figure}

\subsection{Subrelativistic shocks}\label{sec:subR}
At shock velocities $\beta_u <0.3 $ pair creation becomes extremely slow \cite{katz2010,ito2020}, so that practically $q=0$ and $n_+=0$ everywhere inside
the shock.  The electron flux is then constant and equals the proton flux, viz., $n_-u_- = n_pu_p= J_p$.  Moreover, the 
temperature inside and downstream of the shock is well below the electron mass and the pressure is dominated by the radiation.
Thus, we approximate $p_p=p_-=0$ and  $h_- = 1$.  Ignoring the friction term, which is justified by virtue of the minuscule velocity difference found below,
the shock equations simplify to:
\begin{eqnarray}
\frac{d\gamma_p}{dx} = \frac{e}{m_pc^2} E, \label{eq:1D_1b}\\
\frac{d\gamma_-}{dx} = - \frac{e}{m_ec^2} E -\frac{1}{J_p m_ec^2}\frac{dT_r^{0x}}{dx} ,\label{eq:1D_2b}\\
\beta_-^x\frac{dT^{xx}}{dx}= \frac{dT^{x0}}{dx},\label{eq:1D_3b}\\
\frac{dE}{dx}=4\pi e J_p(\beta_p^{-1} - \beta_-^{-1}).\label{eq:1D_4b}
\end{eqnarray}
The above equations must be augmented by some radiative transfer prescription.  A common approach is to employ the diffusion
approximation (see, e.g., Ref. \cite{BP81a}).

As will be shown shortly, the velocity difference between protons and electrons is tiny, $|\beta_p - \beta_-| << \beta_p$. 
Thus, practically $\gamma_p=\gamma_- =\gamma$.  Subtracting Eq. (\ref{eq:1D_1b}) from Eq. (\ref{eq:1D_2b}) 
yields: $e E(1+m_e/m_p) =  - J_p^{-1}dT_r^{0x}/dx \sim - T^{0x}_r/J_p L$, where $L\sim 1/\beta_u n_u\sigma_T$ is the width of the 
shock transition layer.    Approximating $dE/dx \sim E/L = \beta_u n_u \sigma_T E$ we obtain
\begin{equation}
\begin{split}
\beta_-^{-1} - \beta_p^{-1} & \simeq \frac{T_r^{0x}}{4\pi e^2 L^2 J_p^2} \simlt  \frac{\sigma_T^2 m_pc^2 \beta^3_u n_u}{8\pi e^2}\\
&\simeq 10^{-19} \beta^3_u \left(\frac{n_u}{10^{15}\, cm^{-3}}\right),
\end{split}
\end{equation}
where the  inequality is obtained upon assuming that $T^{0x}_{r}$ is smaller than the total upstream energy, viz., $T^{0x}_{r}\simlt J_pm_pc^2(\gamma_u-1)$.
Note that the last equation can be cast in terms of the proton skin depth $l_p$, given by Eq. (\ref{eq:skindepth}) with $\gamma_u=1$,  as: 
\begin{equation}
\beta_-^{-1} - \beta_p^{-1}\simlt \frac{l_p^2}{L^2}\beta_u.
\end{equation}
It is seen that the electrons move slightly slower than the protons to allow generation of electrostatic field that decelerates the protons. 
The same electric field exerts an opposite force on the electrons to counteract the radiative force acting on them.  There is a slight offset, 
by roughly a factor $m_e/m_p$, that results in a net force that gives rise to exactly the same deceleration of electrons  and protons.

With $\gamma_p=\gamma_- =\gamma \approx 1+\beta^2/2$, the sum of Eqs. (\ref{eq:1D_1b}) and (\ref{eq:1D_2b}) 
yields energy conservation equation for the single fluid system,
\begin{equation}
\frac{d}{dx}[ (m_p + m_e) J_p c^2 \beta^2/2 + T_r^{0x}]= 0,
\end{equation}
that together with Eq. (\ref{eq:1D_3b}) and an appropriate  radiative transfer prescription can be solved (see e.g., \cite{weaver1976,BP81b,katz2010}).   
The electric field is obtained from Eq.(\ref{eq:1D_1b}) once $\beta(x)$ is found 
and the velocity difference from Eq.(\ref{eq:1D_3b}) upon substituting $E(x)$.  Numerical integration of Eqs. (\ref{eq:1D_1b})-(\ref{eq:1D_4b})
indicates that small amplitude oscillations are superposed on this analytic solution (see next section).

\subsection{Relativistic pair loaded shocks}\label{sec:RRMS}
At shock velocities $\beta_u \simgt 0.5$ rapid pair creation ensues \cite{ito2020,levinson2020} and the pair density inside the shock transition 
layer and in the immediate downstream largely exceeds the density of baryons.   We shall consider sufficiently relativistic shocks 
($\gamma_u >>1$) for which the two stream approximation, that greatly simplifies the analysis, can be applied to the radiation 
inside the shock transition layer \cite{granot2018,levinson2020}. In this approach one stream consists of back-scattered photons with a net 
density $n'_{\gamma\rightarrow d}$, as measured in the shock frame, that propagate
towards the downstream, while the counterstream contains photons with a density $n'_{\gamma\rightarrow u}$ that were 
generated in the immediate downstream and move towards the upstream.  
The pair production term can be expressed now as 
\begin{eqnarray}
\frac{q}{2} = 2\sigma_{\gamma\gamma}n'_{\gamma\rightarrow u}n'_{\gamma\rightarrow d},
\end{eqnarray}
where $\sigma_{\gamma\gamma}$ is the pair production cross section computed in the momentum frame of 
the photon beam that moves towards the downstream, which in the single fluid approximation assumed to be the local fluid frame.  
In the present case there is no single frame so further considerations are needed (see below).   
Since the mean energy of counterstreaming photons is roughly $m_ec^2$, all scatterings are in the deep Klein-Nishina  (KN)
regime.   Consequently, the energy source term can be approximated as 
\begin{eqnarray}
\frac{S^0_\pm}{m_ec^2} = -2\sigma_{\pm}h_\pm\gamma_\pm n'_\pm n'_{\gamma\rightarrow u} + \frac{q}{2} h_\pm \gamma_\pm,
\end{eqnarray}
where $\sigma_- (\sigma_+)$ is the KN cross section computed in the local momentum frame of the electrons (positron).
The second term on the R.H.S accounts for the net energy added to the
electron and positron fluids through pair creation. 
%

\begin{figure*}[]
\centering
\centerline{ \includegraphics[width=17cm,  keepaspectratio]{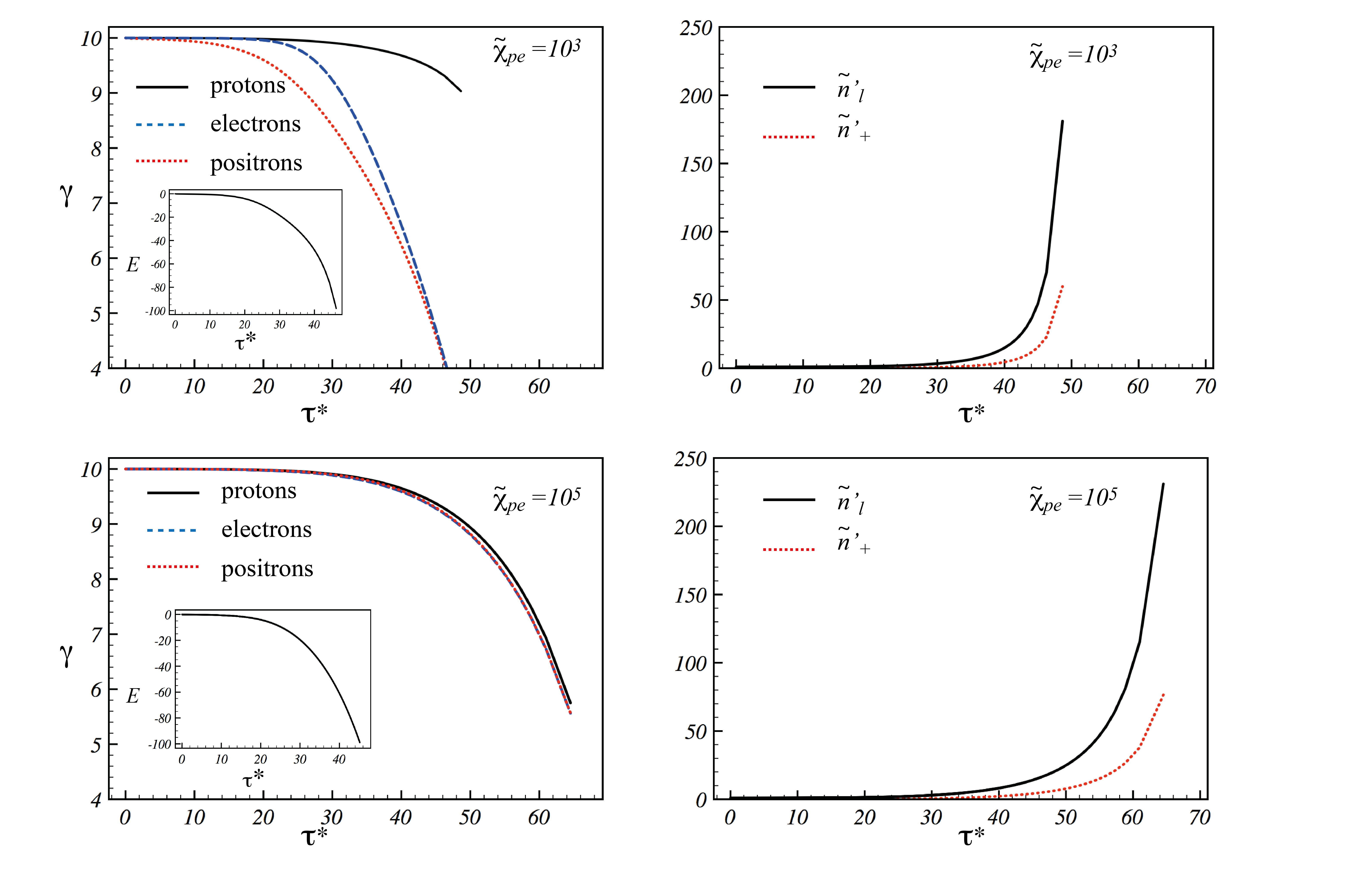}}  
\caption{Solutions obtained for $\tilde{\chi}_{pe}=\tilde{\chi}_{ee} =10^3$ (upper panels), and $10^5$ (lower panels). 
The left panels display the Lorentz factor profiles of the proton, electron and positron fluids, and the right panels the 
positron density and the density of total quanta, $\tilde{n}'_l = 1 +\tilde{n}'_+ + \tilde{n}'_-$. The insets in the left panels
exhibit the electric field $\tilde{E}$.}
\label{fig:fric1}
\end{figure*}

%

The equations are rendered dimensionless by using the coordinate  $d\tau = \sigma_Tn_u\gamma_u dx$,  
and the normalization $\tilde{n}' = n'/\gamma_u n_u = n'/J_p$ for all densities, where $\beta_u=1$ is adopted, 
$\tilde{\sigma} = \sigma/\sigma_T$ for all cross-sections,  and $\tilde{E}=E/E_0$, $\mu=m_e/m_p$.  
The fiducial electric field $E_0$, measured in Gaussian units, and a dimensionless parameter $\chi_E$ to be used below are defined as:
\begin{eqnarray}
E_0 = \frac{m_ec^2 \sigma_T J_p}{e} = 1.2\times10^{-6} n_{u15}\gamma^2_u,\\
\chi_E =\frac{4\pi e}{\sigma_T E_0} = 10^{22} (n_{u15}\gamma_u)^{-1}.
\end{eqnarray}
Note that $\sqrt{\chi_E}/\gamma_u$ is the  ratio of the Thomson length, $\lambda=(\sigma_T n_u\gamma_u)^{-1}$, and the electron skin depth, 
$l_e=c/\omega_e= \sqrt{m_ec^2\gamma_u/(4\pi e^2 n_u)}$.
%
%
In terms of the dimensionless coupling constants $\tilde{\chi}_b=\chi_b/\sigma_Tc$ ($b=pe,ee$), the normalized friction terms 
are given by
\begin{eqnarray}
\tilde{g}_p^0 = -\tilde{\chi}_{pe}[ \tilde{n}'_+ (\gamma_+^{-1} -\gamma_p^{-1})+ \tilde{n}'_- (\gamma_-^{-1} -\gamma_p^{-1})], \label{eq:dim_gp}\\
\tilde{g}_\pm^0 =\pm \tilde{\chi}_{ee} \tilde{n}'_\mp(\gamma_+^{-1} -\gamma_-^{-1})+\tilde{\chi}_{pe} (\gamma_\pm^{-1} -\gamma_p^{-1}).
\end{eqnarray}
The dimensionless coefficients $\tilde{\chi}_a$ can be interpreted as the ratio of the Thomson length and the characteristic 
length over which the momentum of an electron (positron) changes considerably due to collective plasma interactions (see 
Sec. \ref{sec:Disc} for further discussion).

With the above definitions and notations the equations are written in the form:
\begin{eqnarray}
\frac{d \gamma_p}{d\tau} =  \mu \tilde{E} +\mu \tilde{g}_p^0, \label{eq:dg_p}\\
\frac{d (h_\pm \gamma_\pm)}{d\tau} = \pm\, \tilde{E} +\tilde{g}_\pm^0 - 2\tilde{\sigma}_\pm h_\pm\gamma_\pm \tilde{n}' _{\gamma\rightarrow u},\\
\frac{d \tilde{n}' _-}{d\tau} = 2\tilde{\sigma}_{\gamma\gamma} \tilde{n}' _{\gamma\rightarrow u}\tilde{n}' _{\gamma\rightarrow d},\\
\frac{d \tilde{n}' _{\gamma\rightarrow u}}{d\tau} =2 (\tilde{\sigma}_+ \tilde{n}' _+ + \tilde{\sigma}_- \tilde{n}' _- + \tilde{\sigma}_{\gamma\gamma} \tilde{n}' _{\gamma\rightarrow d})\tilde{n}' _{\gamma\rightarrow u},\\
\frac{d \tilde{n}' _{\gamma\rightarrow d}}{d\tau} =2 (\tilde{\sigma}_+ \tilde{n}' _+ + \tilde{\sigma}_- \tilde{n}' _- - \tilde{\sigma}_{\gamma\gamma} \tilde{n}' _{\gamma\rightarrow d})\tilde{n}' _{\gamma\rightarrow u},\\
\frac{d \tilde{E}}{d\tau} =\chi_E \left[\frac{1}{\beta_p} - \frac{1}{\beta_-} + \tilde{n}' _+\beta_+\left(\frac{1}{\beta_+}-\frac{1}{\beta_-}\right)\right], \label{eq:dE}
\end{eqnarray}
with $\tilde{n}' _+ = \tilde{n}' _- -1$, subject to the boundary conditions  $\tilde{n}' _{\gamma\rightarrow u}=\tilde{n}' _{\gamma\rightarrow d}=\tilde{n}' _+ = E =0$  and $\tilde{n}' _- =1$, $\gamma_p=\gamma_- = \gamma_u$ at $\tau=-\infty$.  We also need to specify $h_\pm$ and approximate the cross sections.   One might assume for instance that 
electrons and positrons have the same temperature, so that $h_+=h_- = 4\hat{T}$, and, following Ref \cite{granot2018}, approximate 
\begin{equation}
\hat{T} = \eta \frac{\tilde{n}' _l}{\tilde{n}' _l+2}\frac{\gamma_+ + \gamma_-}{2},\label{eq:Temp}
\end{equation}
where $\tilde{n}' _l=\tilde{n}' _+ + \tilde{n}' _- + \tilde{n}' _{\gamma\rightarrow d}$ is the total density of quanta.  
For the  KN cross sections we adopt the approximation derived in Ref \cite{granot2018}.
\begin{equation}
\tilde{\sigma}_\pm  \approx \frac{3}{8}\cdot\frac{ln(2\gamma_\pm(1+a\hat{{T})})}{\gamma_\pm(1+a\hat{T})}\sigma_{T} .\label{eq:sig}
\end{equation}
Here we use the fact that the energy of counterstreaming photons is $\sim m_e c^2$ with respect to the shock frame, 
and that $\gamma_\pm(1+a\hat{T}) \gg 1$; the factor $a$ accounts for the exact angular distributions of the colliding streams, and
is typically of order unity.   
The pair production cross section has a similar expression with $\gamma_\pm$ replaced by the Lorentz factor $\gamma_r$
associated with the center of momentum frame of the photon beam $n' _{\gamma\rightarrow d}$.  The later can be computed from 
the  energy conservation equation: 
$\gamma_p+\mu h( n' _{+}\gamma_+ + n' _{-}\gamma_- + n' _{\gamma\rightarrow d}\gamma_r)=(1+h_u\mu)\gamma_u$.
To avoid unnecessary complications we approximate  the pair production cross section as
 $\tilde{\sigma}_{\gamma\gamma}=(\tilde{\sigma}_+ +\tilde{\sigma}_-)/2$.    This is accurate enough for our purposes.
Note that in the single fluid approximation,
wherein $\gamma_+=\gamma_-$,  Eqs. (\ref{eq:Temp}) and (\ref{eq:sig}) and  the expression for
$\tilde{\sigma}_{\gamma\gamma}$ reduce to those employed in Ref. \cite{granot2018}.   
Moreover, from Eqs. (\ref{eq:dg_p})-(\ref{eq:dE}) it can be readily shown that this automatically implies $\gamma_r=\gamma_+=\gamma_-$, so
that our model is a direct generalization of the single fluid model developed in Refs. \cite{nakar2012,granot2018}.
Comparison of the single fluid model to numerical simulations \cite{budnik2010,ito2020} indicates an excellent agreement for $\eta$ and $a$ in the range  $0.45-0.55$ and $1.5-2.5$, respectively \cite{granot2018}.   For the calculations presented below we adopted, for illustration, $\eta=0.45$, $a=1.5$.

%

\begin{figure}[]
\centering
\centerline{ \includegraphics[width=8cm]{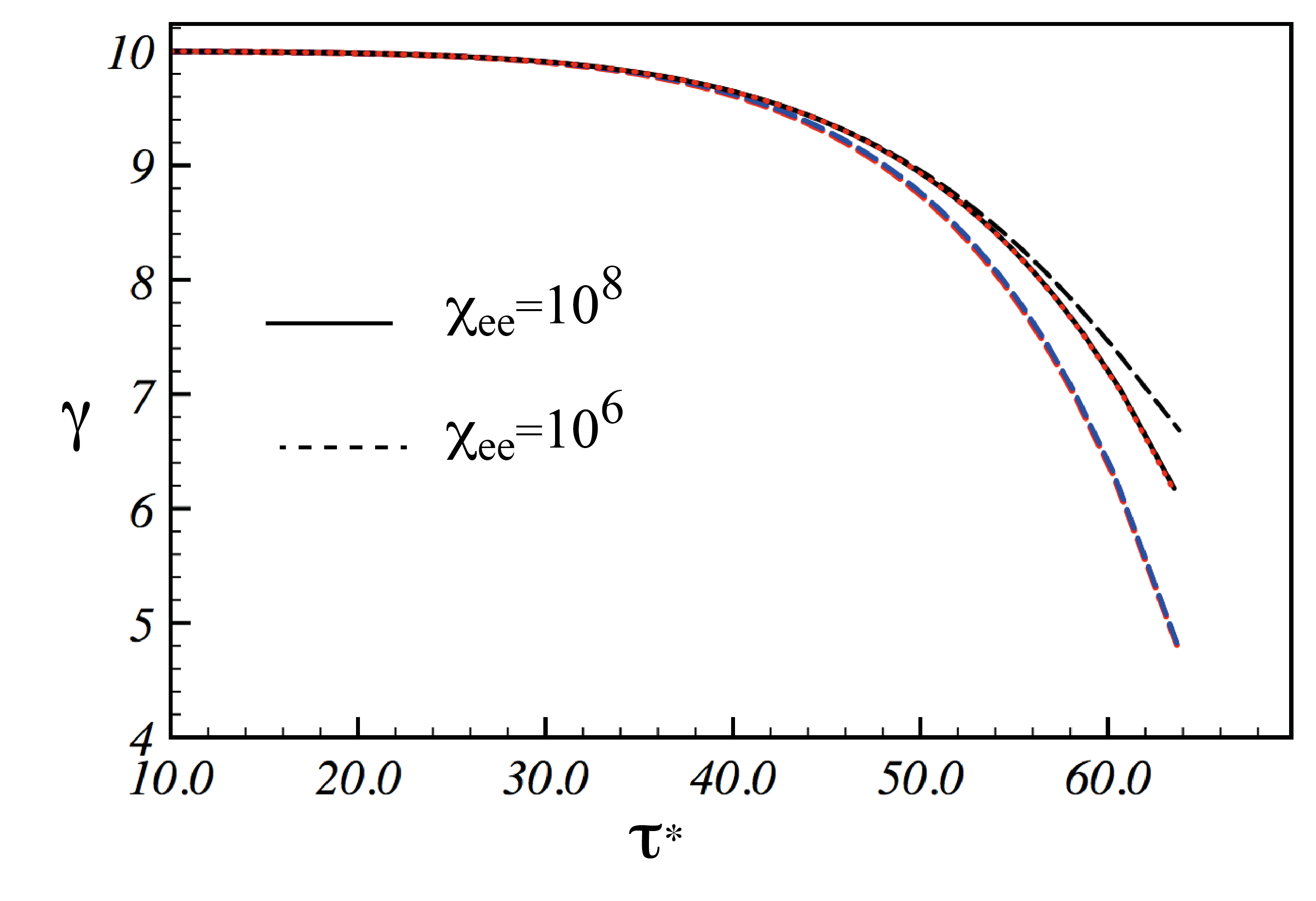}}  
\caption{Lorentz factors profiles of the proton (black lines), electron (blue lines) and positrons (red lines), 
obtained for $\tilde{\chi}_{pe}=0$ and two values of $\tilde{\chi}_{ee}$, $10^6$ (dashed lines) and $10^8$ (solid lines).
The different curves in the latter case are indistinguishable (all seen as the red solid line between the dashed black and 
red/blue lines .}
\label{fig:fric2}
\end{figure}


We solve Eqs. (\ref{eq:dim_gp})-(\ref{eq:dE}) for different values of the parameters $\chi_E$, $\tilde{\chi}_{ee},\tilde{\chi_{pe}}$.    
We do not use realistic values of 
$\chi_E$, as this would require extremely long integration times;  we verified  though that all solutions converge for a large enough value
of $\chi_E$ (between $10^3$ and $10^7$, depending on the specific case), with the exception of the period of small amplitude oscillations of
the electric field, as discussed further below.   To be on the safe side we used $\chi_E=10^8$ in all runs, except for a series of runs aimed
at confirming the anticipated scaling $\sqrt{\chi_E}$ of the oscillations period of $\tilde{E}$, in which we varied $\chi_E$ from $10^4$ to $10^{12}$. 
The integration starts at $\tau=0$, at which we set $\gamma_p=\gamma_\pm=\gamma_u$,
$\tilde{n}'_+=\tilde{n}_{\gamma\rightarrow d} =0$ and $\tilde{n}_{\gamma\rightarrow u}=10^{-2}$.  Choosing a different value for 
$\tilde{n}_{\gamma\rightarrow u}$ merely shifts the origin but does not affect the resultant profiles, provided it is small enough. 

We first examine solutions 
with a negligible friction force, $\chi_{ee}=\chi_{pe}=0$.  An example is shown in Fig. \ref{fig:no_fric} for a shock with an
upstream Lorentz factor $\gamma_u=10$ and $\chi_E=10^7$.  The upper panel exhibits the Lorentz factor profiles of the proton (p), electron ($e^-$)
and positron ($e^+$) fluids, as a function of the pair loaded optical depth, defined in Ref \cite{ito2020}: 
\begin{equation}
\tau^\star=\int \tilde{n}'_l d\tau.
\end{equation}
The evolution of the electric field $\tilde{E}$ is shown in the lower panel.    Superposed on the gradual evolution of $\tilde{E}$  are
small amplitude oscillations (seen more clearly in the inset) with a period equals roughly to the electron skin 
depth $l_e$ (to be precise, the period, when plotted as a function of $\tau$, equals $2\pi\gamma_u^{3/2}/\sqrt{\chi_E}$).    
For realistic values of $\chi_E$ the oscillation period 
will be shorter by about  six orders of magnitude. However, we verified that the Lorentz factor and density profiles are practically independent of 
$\chi_E$ as long as it exceeds $10^3$. 
As seen, initially the positron abundance is minuet and the evolution of the electric field follows roughly that expected in pure electron-proton
flow, as discussed in Sec. \ref{sec:subR} (albeit for a relativistic shock).   During that time the protons and electrons are tightly 
coupled, as seen in the upper panel, but the positrons experience strong deceleration since  in difference from the electrons the electric force
acting on them has the same direction as the radiation force. The increase in positron density (see dashed red line 
in the right panel) leads ultimately to a reversal of the charge polarization  and a consequent  change in the sign of the electric field.
As a result, the electrons decouple from the protons and decelerate fast (roughly over a Thomson length).    In reality, the 
large velocity difference (between any two species) developed at the onset of the shock transition layer should lead to a rapid growth of plasma instabilities,
as observed in collisionless shocks, that may give rise to coupling of the various species via anomalous scattering on plasma turbulence.
A self consistent treatment of plasma kinetic effects requires particle-in-cell simulations of RMS, an extremely challenging problem
given the enormous range of scales involved in this problem, as discussed in the introduction.

Here we investigate the effect of such couplings using our phenomenological prescription of friction.   
We examine two cases; in the first one $\chi_{pe}=\chi_{ee}$, and in second one only the pairs are coupled, that is,
$\chi_{pe}=0$, $\chi_{ee}\ne0$.   Figure \ref{fig:fric1} shows solutions obtained for $\chi_{pe}=\chi_{ee}=10^3$ (upper panles)
and $10^5$ (lower panels).  It is evident that when the friction is strong enough the single fluid model provides a good description
of RMS.   For $\chi_{pe}\approx$ a few times $10^5$ the velocity difference between all species was found to be smaller than the accuracy of 
our calculations.   We compared the Lorentz factor and density profiles obtained for $\chi_{pe}>10^5$ and found a remarkable agreement
with the analytic \cite{katz2010,granot2018,ioka2018} and numerical \cite{budnik2010,ito2020} results obtained for the single fluid model. 

Strong coupling of electrons and positrons alone is, in principle, sufficient to maintain the shock structure close to that of the single fluid model.
In that case, the slight excess of electrons ($\tilde{n}'_- -\tilde{n}'_+=1$) produces the required charge separation in response to the radiation
force, like in a pure $p-e^-$ plasma.   However, we find that a larger value of the dynamical coefficient is needed to tightly couple all 
species compared to the previous case.  Fig. \ref{fig:fric2}  shows the Lorentz factor profiles of $p$ (black lines), $e^+$ (red lines) and 
$e^-$ (blue lines) fluids for $\chi_{pe}=0$ and two values of the pair coupling 
constant, $\tilde{\chi}_{ee}=10^6$ (dashed lines) and $\tilde{\chi}_{ee}=10^8$ (solid lines). 
In the latter case the curves that correspond to the different species are indistinguishable. We find that 
the single fluid model is recovered for $\tilde{\chi}_{ee} \simgt 5\times 10^7$. 

\section{Discussion}\label{sec:Disc}
We have shown that tight coupling of the various plasma constituents (specifically, the proton, electron and positron fluids)
in relativistic radiation mediated shocks requires some form of collective plasma interactions, unlike sub-relativistic shocks
in which minuet charge separation leads to generation of electrostatic field that strongly
couples the ions and electrons.   The absence of sufficient friction between the electron and positron (or proton) fluids 
leads to a large velocity difference already at the onset of the shock transition layer which, as shown by numerical simulations
of collisionless shocks, is prone to various plasma instabilities.    In our analysis, we modelled the collective interactions between 
the fluids by internal friction forces, and quantified the strength of these forces by the dynamical 
coefficients $\chi_{pe}$ and $ \chi_{ee}$. 

The characteristic length scale for momentum transfer from electrons to positrons 
by the friction force, as measured in the shock frame, is related to the dynamical 
coefficient through: $l_{ee} = c/\chi_{ee} n'_-$, and likewise for momentum transfer between electrons and protons.
We can naively interpret this as the mean free path for scattering on plasma waves.  From this we 
deduce that $\tilde{\chi}_{ee} = \chi_{ee}/\sigma_T c = 1/(l_{ee}\sigma_Tn'_-) \approx \lambda/l_{ee}$,
where $\lambda$ is the Thomson length, defined in Eq. (\ref{eq:lambda}) for $\tilde{n}'_-=1$.
In units of the skin depth we have $l_{ee}/l_p = (1/\tilde{\chi}_{ee})(\lambda/l_p)$.  
In the preceding section we found that the values of $\tilde{\chi}_{ee}$ needed for tight coupling are in the range 
$10^5 - 10^8$, depending on details.   By employing Eqs. (\ref{eq:skindepth}) and (\ref{eq:lambda}), 
this can be translated to $l_{ee}/l_p \sim 10^1 - 10^4$ for typical astrophysical  conditions. 

It is unclear at present what is the level of turbulence expected in relativistic RMS.   This issue can only be addressed by 
state-of-the-art particle-in-cell simulations, provided a clever way will be found to rescale the problem, as the huge
dynamic range anticipated by Eqs. (\ref{eq:skindepth}) and (\ref{eq:lambda}) renders full scale simulations infeasible.   
It has been argued (e.g., Refs. \cite{levinson2008,levinson2020}) that particle acceleration is not expected in unmagnetized RMS by 
virtue of this vast separation of scales.   This notion needs to be reconsidered in view of the findings in this paper.  
If second order Fermi acceleration of pairs by the plasma turbulence generated inside the shock can be effective, 
it might  have important implications for detectability of gamma-ray emission from relativistic shock breakouts.

\begin{acknowledgments}
It is a pleasure to thank Evgeny Derishev, Boaz Katz, Yuri Lyubasky, Ehud Nakar and Sasha Philipov for enlightening discussions. 
This research was supported by the  Israel Science Foundation grant 1114/17. 
\end{acknowledgments}


\begin{thebibliography}{30}%
\makeatletter
\providecommand \@ifxundefined [1]{%
 \@ifx{#1\undefined}
}%
\providecommand \@ifnum [1]{%
 \ifnum #1\expandafter \@firstoftwo
 \else \expandafter \@secondoftwo
 \fi
}%
\providecommand \@ifx [1]{%
 \ifx #1\expandafter \@firstoftwo
 \else \expandafter \@secondoftwo
 \fi
}%
\providecommand \natexlab [1]{#1}%
\providecommand \enquote  [1]{``#1''}%
\providecommand \bibnamefont  [1]{#1}%
\providecommand \bibfnamefont [1]{#1}%
\providecommand \citenamefont [1]{#1}%
\providecommand \href@noop [0]{\@secondoftwo}%
\providecommand \href [0]{\begingroup \@sanitize@url \@href}%
\providecommand \@href[1]{\@@startlink{#1}\@@href}%
\providecommand \@@href[1]{\endgroup#1\@@endlink}%
\providecommand \@sanitize@url [0]{\catcode `\\12\catcode `\$12\catcode
  `\&12\catcode `\#12\catcode `\^12\catcode `\_12\catcode `\%12\relax}%
\providecommand \@@startlink[1]{}%
\providecommand \@@endlink[0]{}%
\providecommand \url  [0]{\begingroup\@sanitize@url \@url }%
\providecommand \@url [1]{\endgroup\@href {#1}{\urlprefix }}%
\providecommand \urlprefix  [0]{URL }%
\providecommand \Eprint [0]{\href }%
\providecommand \doibase [0]{http://dx.doi.org/}%
\providecommand \selectlanguage [0]{\@gobble}%
\providecommand \bibinfo  [0]{\@secondoftwo}%
\providecommand \bibfield  [0]{\@secondoftwo}%
\providecommand \translation [1]{[#1]}%
\providecommand \BibitemOpen [0]{}%
\providecommand \bibitemStop [0]{}%
\providecommand \bibitemNoStop [0]{.\EOS\space}%
\providecommand \EOS [0]{\spacefactor3000\relax}%
\providecommand \BibitemShut  [1]{\csname bibitem#1\endcsname}%
\let\auto@bib@innerbib\@empty
\bibitem [{\citenamefont {{Levinson}}\ and\ \citenamefont
  {{Nakar}}(2020)}]{levinson2020}%
  \BibitemOpen
  \bibfield  {author} {\bibinfo {author} {\bibfnamefont {A.}~\bibnamefont
  {{Levinson}}}\ and\ \bibinfo {author} {\bibfnamefont {E.}~\bibnamefont
  {{Nakar}}},\ }\href {\doibase 10.1016/j.physrep.2020.04.003} {\bibfield
  {journal} {\bibinfo  {journal} {\physrep}\ }\textbf {\bibinfo {volume}
  {866}},\ \bibinfo {pages} {1} (\bibinfo {year} {2020})},\ \Eprint
  {http://arxiv.org/abs/1909.10288} {arXiv:1909.10288 [astro-ph.HE]}
  \BibitemShut {NoStop}%
\bibitem [{\citenamefont {{Spitkovsky}}(2008)}]{spitkovsky2008a}%
  \BibitemOpen
  \bibfield  {author} {\bibinfo {author} {\bibfnamefont {A.}~\bibnamefont
  {{Spitkovsky}}},\ }\href {\doibase 10.1086/527374} {\bibfield  {journal}
  {\bibinfo  {journal} {\apjl}\ }\textbf {\bibinfo {volume} {673}},\ \bibinfo
  {pages} {L39} (\bibinfo {year} {2008})},\ \Eprint
  {http://arxiv.org/abs/0706.3126} {arXiv:0706.3126 [astro-ph]} \BibitemShut
  {NoStop}%
\bibitem [{\citenamefont {{Chang}}\ \emph {et~al.}(2008)\citenamefont
  {{Chang}}, \citenamefont {{Spitkovsky}},\ and\ \citenamefont
  {{Arons}}}]{spitkovsky2008b}%
  \BibitemOpen
  \bibfield  {author} {\bibinfo {author} {\bibfnamefont {P.}~\bibnamefont
  {{Chang}}}, \bibinfo {author} {\bibfnamefont {A.}~\bibnamefont
  {{Spitkovsky}}}, \ and\ \bibinfo {author} {\bibfnamefont {J.}~\bibnamefont
  {{Arons}}},\ }\href {\doibase 10.1086/524764} {\bibfield  {journal} {\bibinfo
   {journal} {\apj}\ }\textbf {\bibinfo {volume} {674}},\ \bibinfo {pages}
  {378} (\bibinfo {year} {2008})},\ \Eprint {http://arxiv.org/abs/0704.3832}
  {arXiv:0704.3832 [astro-ph]} \BibitemShut {NoStop}%
\bibitem [{\citenamefont {{Lemoine}}\ and\ \citenamefont
  {{Pelletier}}(2010)}]{lemoine2010}%
  \BibitemOpen
  \bibfield  {author} {\bibinfo {author} {\bibfnamefont {M.}~\bibnamefont
  {{Lemoine}}}\ and\ \bibinfo {author} {\bibfnamefont {G.}~\bibnamefont
  {{Pelletier}}},\ }\href {\doibase 10.1111/j.1365-2966.2009.15869.x}
  {\bibfield  {journal} {\bibinfo  {journal} {\mnras}\ }\textbf {\bibinfo
  {volume} {402}},\ \bibinfo {pages} {321} (\bibinfo {year} {2010})},\ \Eprint
  {http://arxiv.org/abs/0904.2657} {arXiv:0904.2657 [astro-ph.HE]} \BibitemShut
  {NoStop}%
\bibitem [{\citenamefont {{Shaisultanov}}\ \emph {et~al.}(2012)\citenamefont
  {{Shaisultanov}}, \citenamefont {{Lyubarsky}},\ and\ \citenamefont
  {{Eichler}}}]{shaisultanov2012}%
  \BibitemOpen
  \bibfield  {author} {\bibinfo {author} {\bibfnamefont {R.}~\bibnamefont
  {{Shaisultanov}}}, \bibinfo {author} {\bibfnamefont {Y.}~\bibnamefont
  {{Lyubarsky}}}, \ and\ \bibinfo {author} {\bibfnamefont {D.}~\bibnamefont
  {{Eichler}}},\ }\href {\doibase 10.1088/0004-637X/744/2/182} {\bibfield
  {journal} {\bibinfo  {journal} {\apj}\ }\textbf {\bibinfo {volume} {744}},\
  \bibinfo {eid} {182} (\bibinfo {year} {2012})},\ \Eprint
  {http://arxiv.org/abs/1104.0521} {arXiv:1104.0521 [astro-ph.HE]} \BibitemShut
  {NoStop}%
\bibitem [{\citenamefont {{Garasev}}\ and\ \citenamefont
  {{Derishev}}(2016)}]{derishev2016}%
  \BibitemOpen
  \bibfield  {author} {\bibinfo {author} {\bibfnamefont {M.}~\bibnamefont
  {{Garasev}}}\ and\ \bibinfo {author} {\bibfnamefont {E.}~\bibnamefont
  {{Derishev}}},\ }\href {\doibase 10.1093/mnras/stw1345} {\bibfield  {journal}
  {\bibinfo  {journal} {\mnras}\ }\textbf {\bibinfo {volume} {461}},\ \bibinfo
  {pages} {641} (\bibinfo {year} {2016})},\ \Eprint
  {http://arxiv.org/abs/1603.08006} {arXiv:1603.08006 [astro-ph.HE]}
  \BibitemShut {NoStop}%
\bibitem [{\citenamefont {{Lemoine}}\ \emph {et~al.}(2019)\citenamefont
  {{Lemoine}}, \citenamefont {{Gremillet}}, \citenamefont {{Pelletier}},\ and\
  \citenamefont {{Vanthieghem}}}]{lemoine2019}%
  \BibitemOpen
  \bibfield  {author} {\bibinfo {author} {\bibfnamefont {M.}~\bibnamefont
  {{Lemoine}}}, \bibinfo {author} {\bibfnamefont {L.}~\bibnamefont
  {{Gremillet}}}, \bibinfo {author} {\bibfnamefont {G.}~\bibnamefont
  {{Pelletier}}}, \ and\ \bibinfo {author} {\bibfnamefont {A.}~\bibnamefont
  {{Vanthieghem}}},\ }\href {\doibase 10.1103/PhysRevLett.123.035101}
  {\bibfield  {journal} {\bibinfo  {journal} {\prl}\ }\textbf {\bibinfo
  {volume} {123}},\ \bibinfo {eid} {035101} (\bibinfo {year} {2019})},\ \Eprint
  {http://arxiv.org/abs/1907.07595} {arXiv:1907.07595 [astro-ph.HE]}
  \BibitemShut {NoStop}%
\bibitem [{\citenamefont {{Derishev}}\ and\ \citenamefont
  {{Piran}}(2016)}]{derishev2016b}%
  \BibitemOpen
  \bibfield  {author} {\bibinfo {author} {\bibfnamefont {E.~V.}\ \bibnamefont
  {{Derishev}}}\ and\ \bibinfo {author} {\bibfnamefont {T.}~\bibnamefont
  {{Piran}}},\ }\href {\doibase 10.1093/mnras/stw1175} {\bibfield  {journal}
  {\bibinfo  {journal} {\mnras}\ }\textbf {\bibinfo {volume} {460}},\ \bibinfo
  {pages} {2036} (\bibinfo {year} {2016})},\ \Eprint
  {http://arxiv.org/abs/1512.04257} {arXiv:1512.04257 [astro-ph.HE]}
  \BibitemShut {NoStop}%
\bibitem [{\citenamefont {{Levinson}}\ and\ \citenamefont
  {{Bromberg}}(2008)}]{levinson2008}%
  \BibitemOpen
  \bibfield  {author} {\bibinfo {author} {\bibfnamefont {A.}~\bibnamefont
  {{Levinson}}}\ and\ \bibinfo {author} {\bibfnamefont {O.}~\bibnamefont
  {{Bromberg}}},\ }\href {\doibase 10.1103/PhysRevLett.100.131101} {\bibfield
  {journal} {\bibinfo  {journal} {\prl}\ }\textbf {\bibinfo {volume} {100}},\
  \bibinfo {eid} {131101} (\bibinfo {year} {2008})},\ \Eprint
  {http://arxiv.org/abs/0711.3281} {arXiv:0711.3281} \BibitemShut {NoStop}%
\bibitem [{\citenamefont {{Katz}}\ \emph {et~al.}(2010)\citenamefont {{Katz}},
  \citenamefont {{Budnik}},\ and\ \citenamefont {{Waxman}}}]{katz2010}%
  \BibitemOpen
  \bibfield  {author} {\bibinfo {author} {\bibfnamefont {B.}~\bibnamefont
  {{Katz}}}, \bibinfo {author} {\bibfnamefont {R.}~\bibnamefont {{Budnik}}}, \
  and\ \bibinfo {author} {\bibfnamefont {E.}~\bibnamefont {{Waxman}}},\ }\href
  {\doibase 10.1088/0004-637X/716/1/781} {\bibfield  {journal} {\bibinfo
  {journal} {\apj}\ }\textbf {\bibinfo {volume} {716}},\ \bibinfo {pages} {781}
  (\bibinfo {year} {2010})},\ \Eprint {http://arxiv.org/abs/0902.4708}
  {arXiv:0902.4708 [astro-ph.HE]} \BibitemShut {NoStop}%
\bibitem [{\citenamefont {{Nakar}}\ and\ \citenamefont
  {{Sari}}(2010)}]{nakar2010}%
  \BibitemOpen
  \bibfield  {author} {\bibinfo {author} {\bibfnamefont {E.}~\bibnamefont
  {{Nakar}}}\ and\ \bibinfo {author} {\bibfnamefont {R.}~\bibnamefont
  {{Sari}}},\ }\href {\doibase 10.1088/0004-637X/725/1/904} {\bibfield
  {journal} {\bibinfo  {journal} {\apj}\ }\textbf {\bibinfo {volume} {725}},\
  \bibinfo {pages} {904} (\bibinfo {year} {2010})},\ \Eprint
  {http://arxiv.org/abs/1004.2496} {arXiv:1004.2496 [astro-ph.HE]} \BibitemShut
  {NoStop}%
\bibitem [{\citenamefont {{Bromberg}}\ \emph {et~al.}(2011)\citenamefont
  {{Bromberg}}, \citenamefont {{Mikolitzky}},\ and\ \citenamefont
  {{Levinson}}}]{bromberg2011}%
  \BibitemOpen
  \bibfield  {author} {\bibinfo {author} {\bibfnamefont {O.}~\bibnamefont
  {{Bromberg}}}, \bibinfo {author} {\bibfnamefont {Z.}~\bibnamefont
  {{Mikolitzky}}}, \ and\ \bibinfo {author} {\bibfnamefont {A.}~\bibnamefont
  {{Levinson}}},\ }\href {\doibase 10.1088/0004-637X/733/2/85} {\bibfield
  {journal} {\bibinfo  {journal} {\apj}\ }\textbf {\bibinfo {volume} {733}},\
  \bibinfo {eid} {85} (\bibinfo {year} {2011})},\ \Eprint
  {http://arxiv.org/abs/1101.4232} {arXiv:1101.4232 [astro-ph.HE]} \BibitemShut
  {NoStop}%
\bibitem [{\citenamefont {{Sapir}}\ \emph {et~al.}(2011)\citenamefont
  {{Sapir}}, \citenamefont {{Katz}},\ and\ \citenamefont
  {{Waxman}}}]{sapir2011}%
  \BibitemOpen
  \bibfield  {author} {\bibinfo {author} {\bibfnamefont {N.}~\bibnamefont
  {{Sapir}}}, \bibinfo {author} {\bibfnamefont {B.}~\bibnamefont {{Katz}}}, \
  and\ \bibinfo {author} {\bibfnamefont {E.}~\bibnamefont {{Waxman}}},\ }\href
  {\doibase 10.1088/0004-637X/742/1/36} {\bibfield  {journal} {\bibinfo
  {journal} {\apj}\ }\textbf {\bibinfo {volume} {742}},\ \bibinfo {eid} {36}
  (\bibinfo {year} {2011})},\ \Eprint {http://arxiv.org/abs/1103.5075}
  {arXiv:1103.5075 [astro-ph.HE]} \BibitemShut {NoStop}%
\bibitem [{\citenamefont {{Nakar}}\ and\ \citenamefont
  {{Sari}}(2012)}]{nakar2012}%
  \BibitemOpen
  \bibfield  {author} {\bibinfo {author} {\bibfnamefont {E.}~\bibnamefont
  {{Nakar}}}\ and\ \bibinfo {author} {\bibfnamefont {R.}~\bibnamefont
  {{Sari}}},\ }\href {\doibase 10.1088/0004-637X/747/2/88} {\bibfield
  {journal} {\bibinfo  {journal} {\apj}\ }\textbf {\bibinfo {volume} {747}},\
  \bibinfo {eid} {88} (\bibinfo {year} {2012})},\ \Eprint
  {http://arxiv.org/abs/1106.2556} {arXiv:1106.2556 [astro-ph.HE]} \BibitemShut
  {NoStop}%
\bibitem [{\citenamefont {{Sapir}}\ \emph {et~al.}(2013)\citenamefont
  {{Sapir}}, \citenamefont {{Katz}},\ and\ \citenamefont
  {{Waxman}}}]{sapir2013}%
  \BibitemOpen
  \bibfield  {author} {\bibinfo {author} {\bibfnamefont {N.}~\bibnamefont
  {{Sapir}}}, \bibinfo {author} {\bibfnamefont {B.}~\bibnamefont {{Katz}}}, \
  and\ \bibinfo {author} {\bibfnamefont {E.}~\bibnamefont {{Waxman}}},\ }\href
  {\doibase 10.1088/0004-637X/774/1/79} {\bibfield  {journal} {\bibinfo
  {journal} {\apj}\ }\textbf {\bibinfo {volume} {774}},\ \bibinfo {eid} {79}
  (\bibinfo {year} {2013})},\ \Eprint {http://arxiv.org/abs/1304.6428}
  {arXiv:1304.6428 [astro-ph.HE]} \BibitemShut {NoStop}%
\bibitem [{\citenamefont {{Beloborodov}}\ and\ \citenamefont
  {{M{\'e}sz{\'a}ros}}(2017)}]{beloborodov2017b}%
  \BibitemOpen
  \bibfield  {author} {\bibinfo {author} {\bibfnamefont {A.~M.}\ \bibnamefont
  {{Beloborodov}}}\ and\ \bibinfo {author} {\bibfnamefont {P.}~\bibnamefont
  {{M{\'e}sz{\'a}ros}}},\ }\href {\doibase 10.1007/s11214-017-0348-6}
  {\bibfield  {journal} {\bibinfo  {journal} {\ssr}\ }\textbf {\bibinfo
  {volume} {207}},\ \bibinfo {pages} {87} (\bibinfo {year} {2017})},\ \Eprint
  {http://arxiv.org/abs/1701.04523} {arXiv:1701.04523 [astro-ph.HE]}
  \BibitemShut {NoStop}%
\bibitem [{\citenamefont {{Granot}}\ \emph {et~al.}(2018)\citenamefont
  {{Granot}}, \citenamefont {{Nakar}},\ and\ \citenamefont
  {{Levinson}}}]{granot2018}%
  \BibitemOpen
  \bibfield  {author} {\bibinfo {author} {\bibfnamefont {A.}~\bibnamefont
  {{Granot}}}, \bibinfo {author} {\bibfnamefont {E.}~\bibnamefont {{Nakar}}}, \
  and\ \bibinfo {author} {\bibfnamefont {A.}~\bibnamefont {{Levinson}}},\
  }\href {\doibase 10.1093/mnras/sty637} {\bibfield  {journal} {\bibinfo
  {journal} {\mnras}\ }\textbf {\bibinfo {volume} {476}},\ \bibinfo {pages}
  {5453} (\bibinfo {year} {2018})},\ \Eprint {http://arxiv.org/abs/1708.05018}
  {arXiv:1708.05018 [astro-ph.HE]} \BibitemShut {NoStop}%
\bibitem [{\citenamefont {{Ioka}}\ \emph {et~al.}(2019)\citenamefont {{Ioka}},
  \citenamefont {{Levinson}},\ and\ \citenamefont {{Nakar}}}]{ioka2018}%
  \BibitemOpen
  \bibfield  {author} {\bibinfo {author} {\bibfnamefont {K.}~\bibnamefont
  {{Ioka}}}, \bibinfo {author} {\bibfnamefont {A.}~\bibnamefont {{Levinson}}},
  \ and\ \bibinfo {author} {\bibfnamefont {E.}~\bibnamefont {{Nakar}}},\ }\href
  {\doibase 10.1093/mnras/stz270} {\bibfield  {journal} {\bibinfo  {journal}
  {\mnras}\ }\textbf {\bibinfo {volume} {484}},\ \bibinfo {pages} {3502}
  (\bibinfo {year} {2019})},\ \Eprint {http://arxiv.org/abs/1810.11022}
  {arXiv:1810.11022 [astro-ph.HE]} \BibitemShut {NoStop}%
\bibitem [{\citenamefont {{Lundman}}\ \emph {et~al.}(2018)\citenamefont
  {{Lundman}}, \citenamefont {{Beloborodov}},\ and\ \citenamefont
  {{Vurm}}}]{lundman2018a}%
  \BibitemOpen
  \bibfield  {author} {\bibinfo {author} {\bibfnamefont {C.}~\bibnamefont
  {{Lundman}}}, \bibinfo {author} {\bibfnamefont {A.~M.}\ \bibnamefont
  {{Beloborodov}}}, \ and\ \bibinfo {author} {\bibfnamefont {I.}~\bibnamefont
  {{Vurm}}},\ }\href {\doibase 10.3847/1538-4357/aab9b3} {\bibfield  {journal}
  {\bibinfo  {journal} {\apj}\ }\textbf {\bibinfo {volume} {858}},\ \bibinfo
  {eid} {7} (\bibinfo {year} {2018})},\ \Eprint
  {http://arxiv.org/abs/1708.02633} {arXiv:1708.02633 [astro-ph.HE]}
  \BibitemShut {NoStop}%
\bibitem [{\citenamefont {{Lundman}}\ and\ \citenamefont
  {{Beloborodov}}(2019)}]{lundman2018b}%
  \BibitemOpen
  \bibfield  {author} {\bibinfo {author} {\bibfnamefont {C.}~\bibnamefont
  {{Lundman}}}\ and\ \bibinfo {author} {\bibfnamefont {A.~M.}\ \bibnamefont
  {{Beloborodov}}},\ }\href {\doibase 10.3847/1538-4357/ab229f} {\bibfield
  {journal} {\bibinfo  {journal} {\apj}\ }\textbf {\bibinfo {volume} {879}},\
  \bibinfo {eid} {83} (\bibinfo {year} {2019})},\ \Eprint
  {http://arxiv.org/abs/1804.03053} {arXiv:1804.03053 [astro-ph.HE]}
  \BibitemShut {NoStop}%
\bibitem [{\citenamefont {{Lyutikov}}(2018)}]{lyutikov2018}%
  \BibitemOpen
  \bibfield  {author} {\bibinfo {author} {\bibfnamefont {M.}~\bibnamefont
  {{Lyutikov}}},\ }\href {\doibase 10.1093/mnras/sty735} {\bibfield  {journal}
  {\bibinfo  {journal} {\mnras}\ }\textbf {\bibinfo {volume} {477}},\ \bibinfo
  {pages} {816} (\bibinfo {year} {2018})},\ \Eprint
  {http://arxiv.org/abs/1801.04221} {arXiv:1801.04221 [astro-ph.HE]}
  \BibitemShut {NoStop}%
\bibitem [{\citenamefont {{Derishev}}(2018)}]{derishev2018}%
  \BibitemOpen
  \bibfield  {author} {\bibinfo {author} {\bibfnamefont {E.}~\bibnamefont
  {{Derishev}}},\ }\href {\doibase 10.1134/S106377291812020X} {\bibfield
  {journal} {\bibinfo  {journal} {Astronomy Reports}\ }\textbf {\bibinfo
  {volume} {62}},\ \bibinfo {pages} {868} (\bibinfo {year} {2018})},\ \Eprint
  {http://arxiv.org/abs/1812.09866} {arXiv:1812.09866 [astro-ph.HE]}
  \BibitemShut {NoStop}%
\bibitem [{\citenamefont {{Budnik}}\ \emph {et~al.}(2010)\citenamefont
  {{Budnik}}, \citenamefont {{Katz}}, \citenamefont {{Sagiv}},\ and\
  \citenamefont {{Waxman}}}]{budnik2010}%
  \BibitemOpen
  \bibfield  {author} {\bibinfo {author} {\bibfnamefont {R.}~\bibnamefont
  {{Budnik}}}, \bibinfo {author} {\bibfnamefont {B.}~\bibnamefont {{Katz}}},
  \bibinfo {author} {\bibfnamefont {A.}~\bibnamefont {{Sagiv}}}, \ and\
  \bibinfo {author} {\bibfnamefont {E.}~\bibnamefont {{Waxman}}},\ }\href
  {\doibase 10.1088/0004-637X/725/1/63} {\bibfield  {journal} {\bibinfo
  {journal} {\apj}\ }\textbf {\bibinfo {volume} {725}},\ \bibinfo {pages} {63}
  (\bibinfo {year} {2010})},\ \Eprint {http://arxiv.org/abs/1005.0141}
  {arXiv:1005.0141 [astro-ph.HE]} \BibitemShut {NoStop}%
\bibitem [{\citenamefont {{Ito}}\ \emph {et~al.}(2018)\citenamefont {{Ito}},
  \citenamefont {{Levinson}}, \citenamefont {{Stern}},\ and\ \citenamefont
  {{Nagataki}}}]{ito2018a}%
  \BibitemOpen
  \bibfield  {author} {\bibinfo {author} {\bibfnamefont {H.}~\bibnamefont
  {{Ito}}}, \bibinfo {author} {\bibfnamefont {A.}~\bibnamefont {{Levinson}}},
  \bibinfo {author} {\bibfnamefont {B.~E.}\ \bibnamefont {{Stern}}}, \ and\
  \bibinfo {author} {\bibfnamefont {S.}~\bibnamefont {{Nagataki}}},\ }\href
  {\doibase 10.1093/mnras/stx2722} {\bibfield  {journal} {\bibinfo  {journal}
  {\mnras}\ }\textbf {\bibinfo {volume} {474}},\ \bibinfo {pages} {2828}
  (\bibinfo {year} {2018})},\ \Eprint {http://arxiv.org/abs/1709.08955}
  {arXiv:1709.08955 [astro-ph.HE]} \BibitemShut {NoStop}%
\bibitem [{\citenamefont {{Ito}}\ \emph {et~al.}(2020)\citenamefont {{Ito}},
  \citenamefont {{Levinson}},\ and\ \citenamefont {{Nagataki}}}]{ito2020}%
  \BibitemOpen
  \bibfield  {author} {\bibinfo {author} {\bibfnamefont {H.}~\bibnamefont
  {{Ito}}}, \bibinfo {author} {\bibfnamefont {A.}~\bibnamefont {{Levinson}}}, \
  and\ \bibinfo {author} {\bibfnamefont {S.}~\bibnamefont {{Nagataki}}},\
  }\href {\doibase 10.1093/mnras/stz3591} {\bibfield  {journal} {\bibinfo
  {journal} {\mnras}\ }\textbf {\bibinfo {volume} {492}},\ \bibinfo {pages}
  {1902} (\bibinfo {year} {2020})},\ \Eprint {http://arxiv.org/abs/1910.08431}
  {arXiv:1910.08431 [astro-ph.HE]} \BibitemShut {NoStop}%
\bibitem [{\citenamefont {{Beloborodov}}(2017)}]{beloborodov2017a}%
  \BibitemOpen
  \bibfield  {author} {\bibinfo {author} {\bibfnamefont {A.~M.}\ \bibnamefont
  {{Beloborodov}}},\ }\href {\doibase 10.3847/1538-4357/aa5c8c} {\bibfield
  {journal} {\bibinfo  {journal} {\apj}\ }\textbf {\bibinfo {volume} {838}},\
  \bibinfo {eid} {125} (\bibinfo {year} {2017})},\ \Eprint
  {http://arxiv.org/abs/1604.02794} {arXiv:1604.02794 [astro-ph.HE]}
  \BibitemShut {NoStop}%
\bibitem [{\citenamefont {{Zenitani}}\ \emph {et~al.}(2009)\citenamefont
  {{Zenitani}}, \citenamefont {{Hesse}},\ and\ \citenamefont
  {{Klimas}}}]{zenitani2009}%
  \BibitemOpen
  \bibfield  {author} {\bibinfo {author} {\bibfnamefont {S.}~\bibnamefont
  {{Zenitani}}}, \bibinfo {author} {\bibfnamefont {M.}~\bibnamefont {{Hesse}}},
  \ and\ \bibinfo {author} {\bibfnamefont {A.}~\bibnamefont {{Klimas}}},\
  }\href {\doibase 10.1088/0004-637X/696/2/1385} {\bibfield  {journal}
  {\bibinfo  {journal} {\apj}\ }\textbf {\bibinfo {volume} {696}},\ \bibinfo
  {pages} {1385} (\bibinfo {year} {2009})},\ \Eprint
  {http://arxiv.org/abs/0902.2074} {arXiv:0902.2074 [astro-ph.HE]} \BibitemShut
  {NoStop}%
\bibitem [{\citenamefont {{Blandford}}\ and\ \citenamefont
  {{Payne}}(1981{\natexlab{a}})}]{BP81a}%
  \BibitemOpen
  \bibfield  {author} {\bibinfo {author} {\bibfnamefont {R.~D.}\ \bibnamefont
  {{Blandford}}}\ and\ \bibinfo {author} {\bibfnamefont {D.~G.}\ \bibnamefont
  {{Payne}}},\ }\href {\doibase 10.1093/mnras/194.4.1033} {\bibfield  {journal}
  {\bibinfo  {journal} {\mnras}\ }\textbf {\bibinfo {volume} {194}},\ \bibinfo
  {pages} {1033} (\bibinfo {year} {1981}{\natexlab{a}})}\BibitemShut {NoStop}%
\bibitem [{\citenamefont {{Weaver}}(1976)}]{weaver1976}%
  \BibitemOpen
  \bibfield  {author} {\bibinfo {author} {\bibfnamefont {T.~A.}\ \bibnamefont
  {{Weaver}}},\ }\href {\doibase 10.1086/190398} {\bibfield  {journal}
  {\bibinfo  {journal} {\apjs}\ }\textbf {\bibinfo {volume} {32}},\ \bibinfo
  {pages} {233} (\bibinfo {year} {1976})}\BibitemShut {NoStop}%
\bibitem [{\citenamefont {{Blandford}}\ and\ \citenamefont
  {{Payne}}(1981{\natexlab{b}})}]{BP81b}%
  \BibitemOpen
  \bibfield  {author} {\bibinfo {author} {\bibfnamefont {R.~D.}\ \bibnamefont
  {{Blandford}}}\ and\ \bibinfo {author} {\bibfnamefont {D.~G.}\ \bibnamefont
  {{Payne}}},\ }\href {\doibase 10.1093/mnras/194.4.1041} {\bibfield  {journal}
  {\bibinfo  {journal} {\mnras}\ }\textbf {\bibinfo {volume} {194}},\ \bibinfo
  {pages} {1041} (\bibinfo {year} {1981}{\natexlab{b}})}\BibitemShut {NoStop}%
\end{thebibliography}
%
%

\end{document}